\documentclass[twocolumn,english,prb]{revtex4}
\usepackage[T1]{fontenc}
\usepackage[latin9]{inputenc}
\usepackage{amsmath}
\usepackage{graphicx}

\makeatletter
\@ifundefined{textcolor}{}
{%
 \definecolor{BLACK}{gray}{0}
 \definecolor{WHITE}{gray}{1}
 \definecolor{RED}{rgb}{1,0,0}
 \definecolor{GREEN}{rgb}{0,1,0}
 \definecolor{BLUE}{rgb}{0,0,1}
 \definecolor{CYAN}{cmyk}{1,0,0,0}
 \definecolor{MAGENTA}{cmyk}{0,1,0,0}
 \definecolor{YELLOW}{cmyk}{0,0,1,0}
}

\DeclareMathSizes{10}{10}{7}{5}

\usepackage{babel}

\makeatother

\usepackage{babel}
\begin{document}

\title{local probing of nuclear bath polarization with a single electronic
spin}

\author{P. London,$^{1,*}$ R. Fischer,$^{1}$ I. Alvizu,$^{2}$ J. R. Maze,$^{2}$
and D. Gershoni$^{1}$}

\affiliation{$^{1}$Department of Physics, Technion, Israel Institute of Technology,
Haifa 32000, Israel}

\affiliation{$^{2}$Departmento de Fisica, Pontificia Universidad Catolica de
Chile, Santiago 7820436, Chile}
\begin{abstract}
We demonstrate experimentally that a polarized nuclear spin modifies
the dynamic behavior of a neighboring electronic spin. Specifically,
an out-of-phase component appears in the electronic spin-echo signal.
This component is proportional to the nuclear spin degree of polarization
and strongly depends on the nuclear polarization direction. When the
electronic spin is surrounded by a polarized nuclear spin bath, the
spin-echo quadrature manifests a characteristic frequency related
only to the nuclear spins abundance and their collective polarization.
We use this analysis to propose a novel measurement method for the
local nuclear spin bath of a single electronic spin. We quantify the
realistic experimental regimes at which the scheme is efficient. Our
proposal has potential applications for quantum sensing schemes, and
opens a route for a systematic study of polarized mesoscopical-systems.
\end{abstract}
\maketitle
Enhancement of nuclear polarization via polarization transfer from
electronic spins is a basic ingredient in nuclear magnetic resonance
(NMR) science, and a promising approach for enhancing the sensitivity
of nuclear spin based applications, such as magnetic resonance imaging
(MRI). For quantum information processing (QIP) and quantum meterology
studies, nuclear bath polarization is essential for initializing the
state of the system, for instance in a quantum simulator\cite{CaiNatPhys2013},
or for increasing the signal-to-noise ratio (SNR)\cite{MaminScience2013,StaudacherScience2013,AjoyPRX2015}.
However, measuring the polarization of nuclear spins is a challenge
due to their tiny magnetic moment. Possible solutions tackling this
difficulty are measurements involving large ensembles or the search
for electronic spins energy shifts due to static nuclear spin polarization
\cite{KornackPRL2005,HapperRevModPhys1997,MakhoninNatMat2011}. 

An additional approach is to probe the influence of the nuclear polarization
on the \textbf{dynamical} behavior of a central electronic spin \cite{LondonPRL2013,LiuNanoLett2014}.
The latter is most appealing since it relays on the same mechanism
that governs the electron-nuclear polarization\cite{AlvarezArXiv2014,TaminiauNatNano2014}.
Specifically, dynamical decoupling protocols suppress the noise originating
from the nuclear-spin bath\cite{DeLangeScience2010,TaylorNatPhys2008,NaydenovPRB2011}.
However, they also dictate immunity to any static field originating
from the surrounding bath, thus preventing its measurement. Surprisingly,
when the central spin coherently interacts with the surrounding bath,
the mutual dynamics of the spin and its environment can provide useful
information on the bath polarization, and the bath polarization can
actually be used for improving the spin sensitivity to magnetic fields\cite{GoldsteinPRL2011}.

Here, we analyze the effect of nuclear bath polarization on a prototypical
central spin system - the nitrogen-vacancy (NV) color center in diamond
interacting with a bath of $^{13}$C nuclear spins. The NV-center
in diamond is a promising physical platform for QIP and nanoscale
metrology; its ground state sub-levels are optically accessible, can
be coherently manipulated using MW fields, and present unprecedentedly
long coherence times for a solid state system at room-temperature
\cite{GopiNatMat2009}. These properties have engaged a number of
important NV-center demonstrations in QIP \cite{WaldherrNature2014},
nanoscale magnetometry \cite{GopiNature2008,MazeNature2008}, nanoscale
NMR \cite{MaminScience2013,StaudacherScience2013}, and measurements
in living cells\cite{KucskoNature2013,McGuinnessNnano2011}. Here,
we study the dynamics of the NV-center interacting with a polarized
nuclear environment under the simplest, yet powerful, dynamical decoupling
protocol - the spin-echo sequence\cite{HahnPR1950}. 

An illustration of the model system is given in Fig.1a; It comprises
the electronic spin of an NV-center , and an ensemble of nuclear spins
randomly distributed in a diamond lattice in the presence of an external
magnetic field $\mathbf{B}$. The experimental pulse sequence is given
in Fig.1b. In spin-echo measurement a $\left(\pi/2\right)$ pulse
rotates the initialized state $\left|0\right\rangle $ to a $\left(1/\sqrt{2}\right)\left[\left|0\right\rangle +\left|1\right\rangle \right]$
superposition, $\left|0\right\rangle ,\left|1\right\rangle $ being
the eigenstates of the electronic spin. This superposition accumulates
dynamical phase according to the local magnetic field at the electronic-spin
position\cite{Slichter}, but tends to decohere after a short time,
$T_{2}^{\star}$. An additional $\pi$ pulse after a duration $\tau$
will result with a revival of the electronic coherence, $S$, after
an identical duration\cite{HahnPR1950}. In an NV-based measurement,
an additional $\left(\pi/2\right)$ pulse rotates the electronic coherence
into a measurable population difference of the ground state sublevels,
detectable by a short optical pulse. Reconstruction of the magnitude
\textbf{and} phase of the coherence ($\mathit{Quadrature}$ $\mathit{detection}$)
is achieved by extracting the real component from an in-phase ($I$)
pulses sequence $\left(\frac{\pi}{2}\right)_{x}\underset{\tau}{-}\left(\pi\right)_{x}\underset{\tau}{-}\left(\frac{\pi}{2}\right)_{x}$,
and the imaginary component from the out-of-phase ($Q$) sequence
$\left(\frac{\pi}{2}\right)_{x}\underset{\tau}{-}\left(\pi\right)_{x}\underset{\tau}{-}\left(\frac{\pi}{2}\right)_{\text{y}}$
\cite{Comment2,SiForBathPolarizationPaper}(Fig. 1b). The coherence
$S$, also referred to as the pseudo spin, can be expressed using
density matrix formalism as $S=\prod_{k}S_{k}$, where \cite{MimsPRB1972,RowanPR1965,MazePRB2008,KolkowitzPRL2012}
\begin{equation}
S_{k}=\mathrm{Tr}_{nuc}\left(U_{1}^{k\dagger}U_{0}^{k\dagger}U_{1}^{k}U_{0}^{k}\rho_{k}\right).\label{eq:SpmTerm}
\end{equation}
Here, $U_{m_{s}}^{k}=\exp$$\left(-iH_{m_{s}}^{k}\tau\right)$ represents
the evolution operator of the k-th nuclear spin conditioned by the
electron spin state $m_{s}=0\left(1\right)$, and $H_{m_{s}}^{k}=\frac{\omega_{m_{s}}^{k}}{2}\mathbf{\vec{\sigma}^{\mathit{k}}}\cdot\widehat{\mathbf{n}}_{m_{s}}^{k}$
is its corresponding Hamiltonian, where $\vec{\sigma}^{k}$ are the
Pauli matrices vector of the k-th nuclear spin, and $\omega_{m_{s}}^{k}\widehat{\mathbf{n}}_{m_{s}}^{k}=\gamma_{n}\mathbf{B}+m_{s}\mathbf{A_{\mathit{k}}}$.
Here, the vector $\mathbf{A_{\mathit{k}}}$ characterizes the interaction
between the k-th nuclear spin and the electronic spin under the secular
approximation. Finally, $\rho_{k}$ in Eq. (\ref{eq:SpmTerm}) is
a density matrix characterizing the initial state of the k-th nuclear
spin. We note that $\omega_{m_{s}=0}$ and $\widehat{\mathbf{n}}_{m_{s}=0}$
are common to all nuclear spins.

The dynamics reflected from Eq.(\ref{eq:SpmTerm}) was previously
considered and measured \cite{ZhaoNnano2012,TaminiauPRL2012,KolkowitzPRL2012},
under the assumption that the nuclear spin-bath is unpolarized, $\rho_{k}=\frac{1}{2}\mathbf{1}$
(${\it high}$ $temperature$ $limit$). The focus of this work is
to introduce polarization to the nuclear system, and to investigate
its influence on the dynamical behavior of the electronic spin NV-center.
We introduce the bath nuclear polarization as a non-coherent state\cite{CommentForBathPOlarizationPaper},
$\prod_{k}\rho_{k}$ , where $\rho_{k}=\frac{1}{2}\mathbf{1}+\frac{P_{k}}{2}\mathbf{\sigma^{\mathit{k}}}\cdot\widehat{\mathbf{m}}^{k}$,
$P_{k}$ being the initial degree-of-polarization of the k-th spin
($-1\leq P_{k}\leq1$), and $\widehat{\mathbf{m}}^{k}$ is the polarization
direction.

For a single (k-th) nuclear spin, Eq.(\ref{eq:SpmTerm}) can be expressed
explicitly. Assuming that the nuclear polarization is oriented along
the external magnetic field axis ($\widehat{\mathbf{m}}^{k}=\widehat{\mathbf{n}}_{0}$),
we find 
\begin{eqnarray}
S_{k} & = & 1-\left|\widehat{\mathbf{n}}_{0}\times\widehat{\mathbf{n}}_{1}^{k}\right|^{2}\sin^{2}\left(\frac{\omega_{1}^{k}\tau}{2}\right)\left(R_{k}e^{i\Theta_{k}}-1\right)\label{eq:PseudoSpinSingleNuc}
\end{eqnarray}
where $R_{k}e^{i\Theta_{k}}$ is the polar representation of $\cos\left(\omega_{0}\tau\right)-iP_{k}\sin\left(\omega_{0}\tau\right)$.
The spin-echo envelope modulation formula \cite{RowanPR1965,ChildressScience2006}
is given by the real part of Eq. (\ref{eq:PseudoSpinSingleNuc}),
and is independent of the nuclear spin state. In contrast, the imaginary
part of $S_{k}$ is proportional to the polarization \textbf{$P_{k}$}.
Fig.1c, and Fig.1d depict the temporal evolution of both $S_{k}$
components for the unpolarized and polarized case, respectively. The
insets to Fig.1(c,d) present the trajectory of $S_{k}$ in the complex
plane.

\begin{figure}
\includegraphics[bb=0bp 30bp 730bp 550bp,clip,width=1\columnwidth]{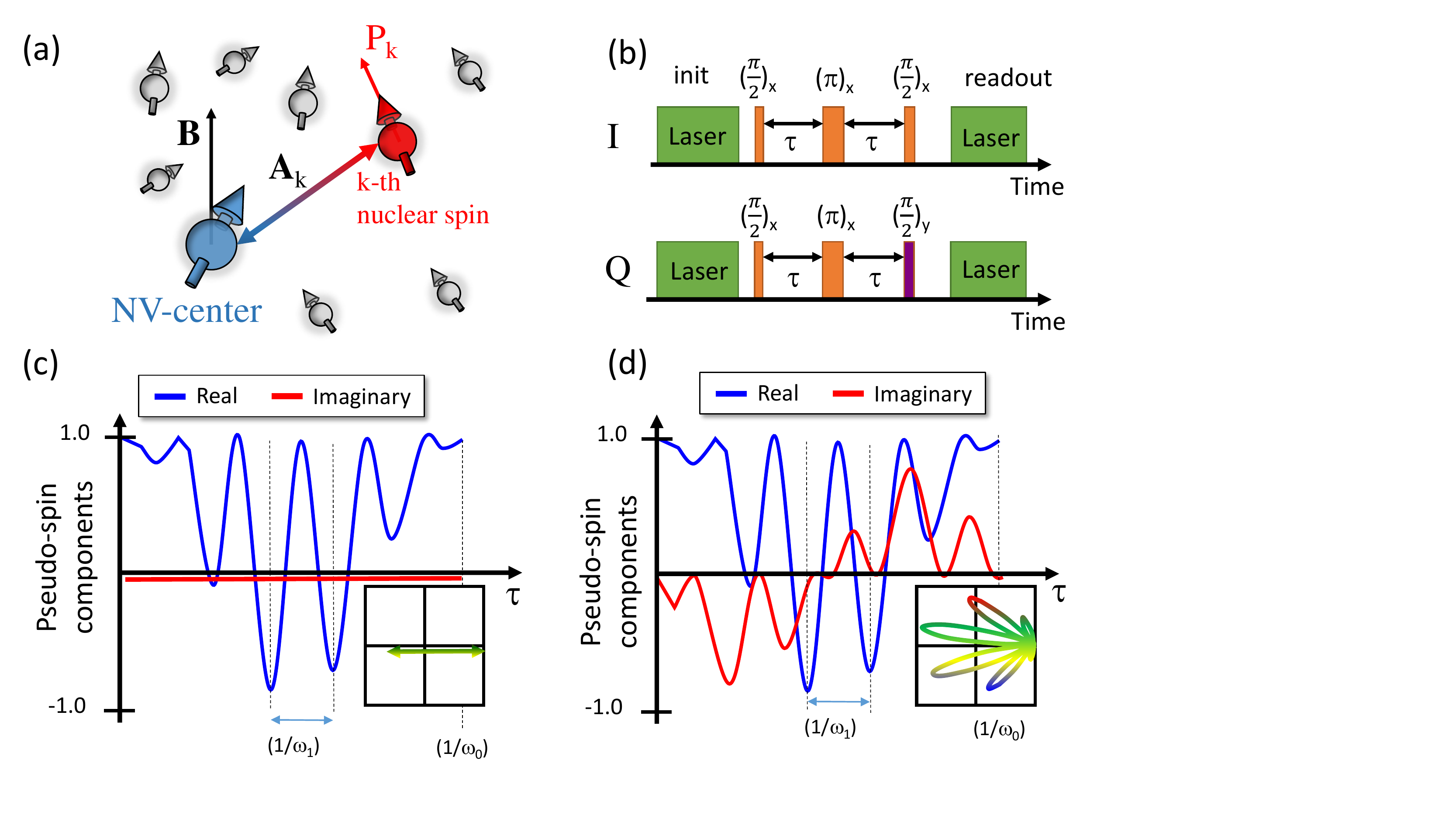} 

\protect\protect\caption{Spin-echo measurement in the presence of a polarized nuclear spin.
(a) A single electronic spin interacts with a bath of nuclear spins
subject to an external magnetic field (b) Schematic description of
the spin-echo pulse sequence. I - in-phase sequence, Q - out-of-phase
sequence. (c){[}(d){]} The pseudo spin ($S$) components as calculated
from Eq.(\ref{eq:PseudoSpinSingleNuc}) using $\omega_{1}=6\omega_{0}$
for unpolarized {[}polarized{]} nuclear spin ; The real value is obtained
by the in-phase sequence (I), and the imaginary values are obtained
by the out-of-phase sequence (Q). The insets to Fig.1(c,d) present
the trajectory of $S_{k}$ in the complex plane.}
\end{figure}

To validate the predictions of our theory, we performed experiments
with a single NV center interacting with a single $^{13}$C whose
polarization is controlled at will. The system is represented with
the electron spin states $\left|0\right\rangle ,\left|-1\right\rangle $
and with the nuclear spin states $\left|\alpha_{\pm}\right\rangle $
which are the eigen-states of $H_{m_{s}=1}$ (Fig. 2a). It has a characteristic
splitting $\Delta=\left(2\pi\right)9$MHz between the nuclear states
within the $\left|1\right\rangle $ manifold, and rotation frequency
$\delta=\left(2\pi\right)0.06$MHz between the nuclear states within
the $\left|0\right\rangle $ manifold (determined by our magnetic
field alignment). Fig. 1b schematically describes the three principle
steps in the experiment: A long laser pulse polarizes the electron
spin and depolarizes the nuclear spin \cite{JiangPRL2008} (step 1).
Then, MW and optical pumping operations are synchronized with the
rotation $\delta$ to efficiently polarize the nuclear spin to one
of the $\left|\alpha_{\pm}\right\rangle $ \cite{DuttScience2007}
(step 2, for details on our experimental parameters see \cite{SiForBathPolarizationPaper}).
Finally, the $I$ or $Q$ echo sequences are employed, and are followed
by a readout laser (step 3). Fig. 2c presents the measured spin-echo
$I$,$Q$ signals, when the nuclear spin was either polarized or remained
unpolarized (denoted pol and ref, respectively). The collapse of electron
spin coherence is accompanied with a fast modulation at $\Delta$
frequency, as predicted by Eq. (\ref{eq:PseudoSpinSingleNuc}) (Fig2c,
$I$-signals). The same frequency appears in the $Q$-signal only
if the nuclear spin is initially polarized. Fourier analysis quantifies
the amplitude of this modulation (Fig.2d, starred peak) enabling a
nuclear polarization quantitative estimation. For this strongly coupled
spin, one has an direct measurement of the nuclear polarization: during
free evolution, the nuclear state precesses between the $\left|\alpha_{\pm}\right\rangle $
states periodically \cite{ShimArXiv2013}. This precession can be
observed with a MW1 $\pi$-pulse and laser readout, and its amplitude
is proportional to the nuclear polarization, $P_{k}$. In our experiments,
we polarized the nuclear spin to its $\left|\alpha_{-}\right\rangle $
state and measured the nuclear polarization using both techniques,
i.e. our quadrature spin-echo technique and the direct method. The
former forms the y-axis and the latter forms the x-axis in Fig. 2e.
(The starred point in Fig.2e represents the data extracted from the
Q-signal curve in Fig.2d, which is marked by a star). When a laser
pulse of various durations was applied between the nuclear polarization
step and the nuclear polarization measurement step (Fig.2e, inset),
we have observed a gradual decrease in the nuclear polarization \cite{JiangPRL2008},
and established the $\mathit{linear}$ dependence of our signal to
the polarization degree $P_{k}$ (Fig.2e). Moreover, we have used
the precession between the $\left|\alpha_{\pm}\right\rangle $ states
to characterize the dependence of the Q-signal in the polarization
direction $\widehat{\mathbf{m}}^{k}$ (Fig.2f, inset). Performing
the quadrature detection at various times, we find strong modulation
of the Q-signal as the nuclear spin rotates prior to the spin-echo
measurement (Fig.2f). Numerical propagation of Eq.(\ref{eq:SpmTerm})
reproduces these results, when the initial nuclear density matrix
$\rho_{k}$ is introduced to the simulation according to free precession
of the $\left|\alpha_{-}\right\rangle $ state around $\mathbf{B}$
(Fig. 2f, red line).

\begin{figure}[b]
\includegraphics[clip,width=1\columnwidth]{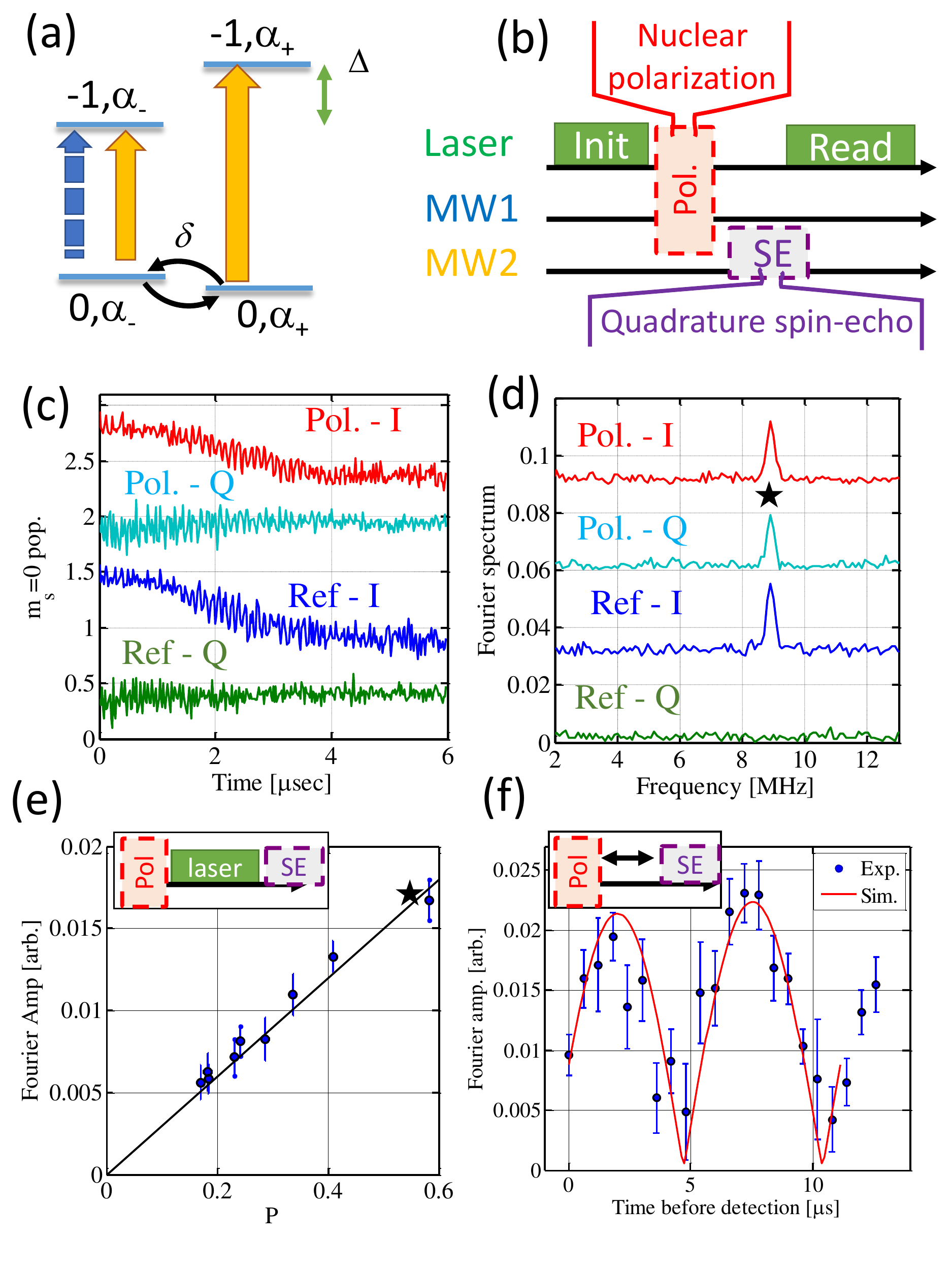}

\protect\caption{Experimental demonstration of nuclear polarization effect on the spin-echo
signal. (a) Energy levels and states of the interacting electronic
($\left|0\right\rangle ,\left|-1\right\rangle $) and nuclear ($\left|\alpha_{\pm}\right\rangle $)
spins. The dashed blue arrow represents resonant MW field which acts
on the $\left|0\alpha_{-}\right\rangle $ level only, while the two
solid orange arrows represent MW field which acts on both levels.
(b) Schematic description of the experimental steps (see main text).
(c) Measured quadrature spin-echo signals (Pol. - polarized nuclear
spin, Ref. - unpolarized nuclear spin). (d) The Fourier transforms
of the signals in (c). Note that the $\omega_{1}$ resonance in the
Q-signal (marked by a star) appears in the polarized case only. (e)
The amplitude of the $\omega_{1}$ resonance in the Q-signal as a
function of the nuclear polarization. The inset illustrates the addition
of a laser pulse used to vary the degree of nuclear polarization.
the data point marked by star corresponds to the measurement in (d).
(f) The $\omega_{1}$ resonance in the Q-signal as a function of the
temporal delay between the nuclear polarization step and spin-echo
measurement, as schematically described in the inset. The solid red
curve represent theoretical simulations (see text).}
 
\end{figure}

We now show that the polarization of a nuclear spin ${\it bath}$
in the NV-center surrounding can be extracted from this protocol.
As each of the $S_{k}$ terms in Eq.(\ref{eq:SpmTerm}) is a complex
number, the calculation of the total pseudo-spin $S$ is merely a
multiplication of their amplitude and a summation of their phase.
It gives $S=\Lambda\left(t\right)e^{i\Phi\left(t\right)}$, where
$\Lambda\left(t\right)$ has the ``collapse and revival'' character
and 
\begin{align}
\Phi\left(\tau\right) & =\sum_{k}\tan^{-1}\left(P_{k}\frac{C_{0}}{S_{0}}\frac{2\left|\widehat{\mathbf{n}}_{0}\times\widehat{\mathbf{n}}_{1}^{k}\right|^{2}S_{0}^{2}\left(S_{1}^{k}\right)^{2}}{2\left|\widehat{\mathbf{n}}_{0}\times\widehat{\mathbf{n}}_{1}^{k}\right|^{2}S_{0}^{2}\left(S_{1}^{k}\right)^{2}-1}\right).\label{eq:ExactPseudoSpin}
\end{align}
Here, $S_{0\left(1\right)}=\sin\left(\frac{\omega_{0\left(1\right)}\tau}{2}\right)$
and $C_{0\left(1\right)}=\cos\left(\frac{\omega_{0\left(1\right)}\tau}{2}\right)$.
Importantly, though each nuclear spin possesses only a small imaginary
term, the total angle, being the sum of many nuclear spins, can be
finite. This leads to the characteristic behavior illustrated in Fig.3a.
Here, the oscillations of the $S$ components (essentially, a rotation
of $S$ in the complex plane) are seen at the revival times. In contrast
to the single nuclear spin case (Fig.1c,d), in the polarized bath
case the $I$-signal is modified by the polarized nuclear bath, in
addition to the dramatic change in the $Q$-signal. At the revival
times ($\omega_{0}t_{r}\simeq2\pi$), the phase accumulation \textbf{rate}
can be approximated. It is 
\begin{eqnarray}
\varpi=\left.\frac{d\Phi\left(t\right)}{dt}\right|_{t_{r}} & \simeq & -\frac{\omega_{0}}{2}\sum_{k}\left|\widehat{\mathbf{n}}_{0}\times\widehat{\mathbf{n}}_{1}^{k}\right|^{2}P_{k},\label{eq:RevivalApproximate}
\end{eqnarray}
and it articulates that at the revival times, $\varpi$ correlates
with the total magnetization in the NV-center surroundings. The weighting
factor $\left|\widehat{\mathbf{n}}_{0}\times\widehat{\mathbf{n}}_{1}^{k}\right|^{2}$
ensures convergence of the sum, and expresses the importance of nearby
nuclear spins (alternatively quantify at which magnetic field one
should expect a prominent signal) \cite{ReinhardPRL2012}. Therefore,
we propose to use $\varpi$ is a quantitative measure for the effective
magnetization in the NV-center vicinity. Since the oscillations are
only observed during the revival time of the spin-echo modulation,
the revival duration $\Delta T$ influences the oscillations contrast
$C$ roughly as $C\simeq\exp\left[-\left(\frac{\pi}{\varpi\Delta T}\right)^{2}\right]$,
and determines a lower limit for the detectable magnetization. 

In our simulations, nuclear spins were randomly positioned in their
lattice sites yielding a desired $^{13}$C abundance. A hollow-sphere
configuration was used ($0.65$nm$\leq R\leq5.5$nm) for omitting
the strongly coupled nuclear spins; these spins are not described
adequately by the dipole term taken in Eq.(\ref{eq:SpmTerm}), since
their hyperfine interaction mixes the electron and nuclear states
\cite{ChildressScience2006}. Moreover, these spins modify the signal
significantly, and obscure the universal behavior of an NV-center
surrounded by a polarized bath. Finally, the polarization of these
strongly coupled spins can be monitored directly, for example using
the method described in \cite{DuttScience2007,SiForBathPolarizationPaper}
Fig. 3b shows the simulated Q-signals at a magnetic field of $B=10$
G and natural $^{13}$C abundance ($n=0.01$). Nuclear polarizations
at the level of \textasciitilde{}tens of percents produce a measurable
$Q$ (Contrast values are stated to the right of each signal). The
inset manifests the linear dependence of $\varpi$ in the polarization
degree {[}blue squares are best fitted values to a sine function with
a Gaussian envelope, and red solid line is the prediction of Eq.(\ref{eq:RevivalApproximate}){]},
and emphasizes that $\varpi$ could serve as a bath polarization probe.
Fig.3c summarize the influence of the physical regime (magnetic fields
and $^{13}$C abundances) on the observable $\varpi$, and essentially
illustrate the dependence of this phenomenon in the pre-factors $\left|\widehat{\mathbf{n}}_{0}\times\widehat{\mathbf{n}}_{1}^{k}\right|^{2}$.
The total effect is quenched by the increasing magnetic field, but
could be recovered by increasing the number of contributing nuclear
spins. Fig.3d supplies the contrast $C$ which is required for evaluating
the scheme's efficiency at the various regimes. Observable signal
is expected at relatively low magnetic fields $B\leq50$G, even for
diamonds with natural $^{13}$C abundance (for example, $\varpi$=50kHz
and $C$=90\%, at $B=5$ G, $n=0.01$). At magnetic fields of $B=500$
G, $\widehat{\mathbf{n}}_{0}\times\widehat{\mathbf{n}}_{1}^{k}$ are
relatively small and collapse-and-revival features are not observed
for natural abundance diamonds. Accordingly, the oscillatory behavior
vanishes. For higher $^{13}$C concentrations, however, the expected
contrast is $C\sim10$\%, and the corresponding frequency is $\varpi\sim10$
kHz. The latter regime is specifically interesting because it promotes
nuclear-bath polarization through excited-state level anti-crossing
method \cite{JacquesPRL2009,FischerPRB2013,FischerPRL2013}).

\begin{figure}
\includegraphics[bb=10bp 0bp 670bp 540bp,clip,width=1\columnwidth]{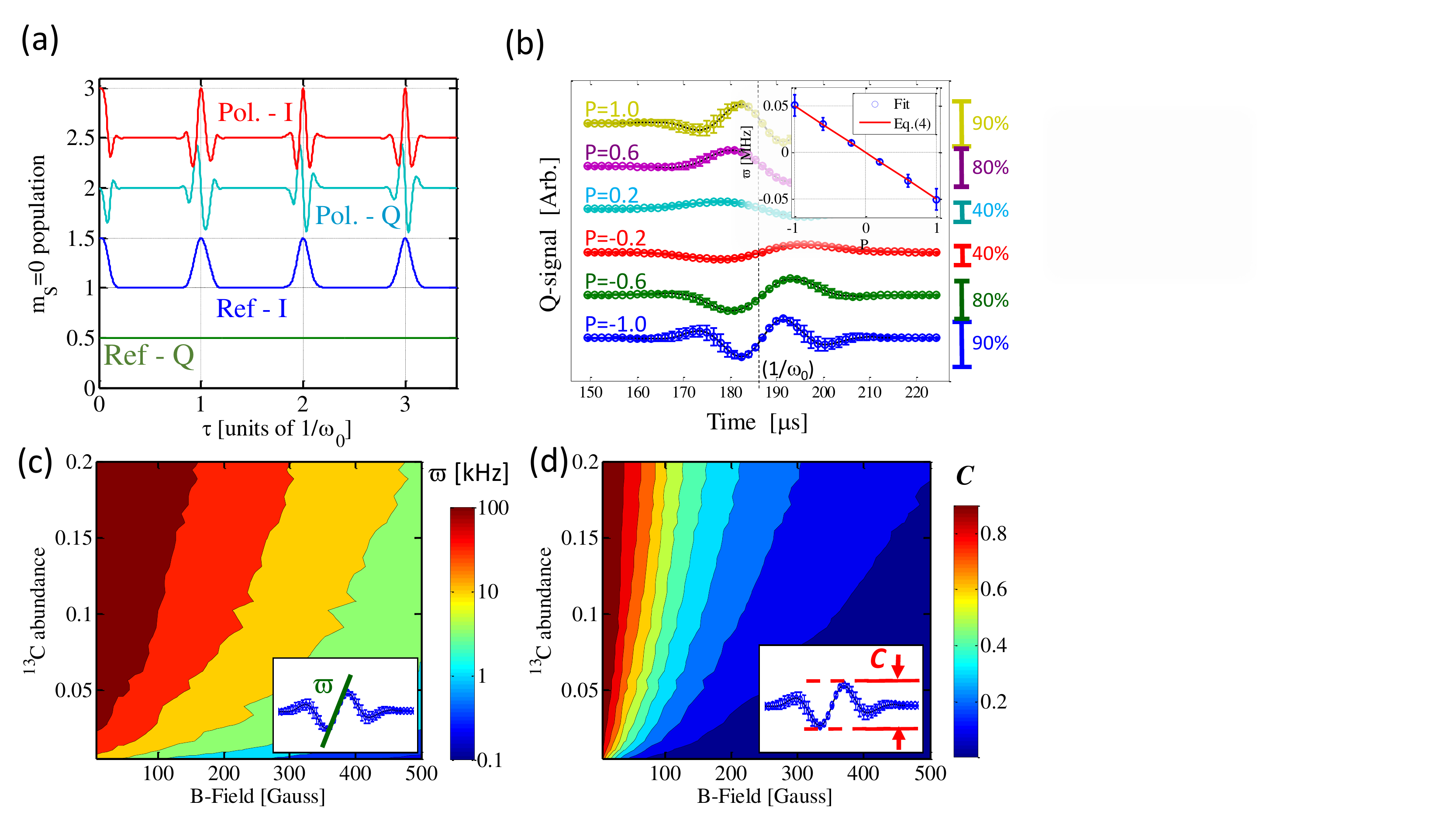}

\protect\protect\caption{Characteristic frequencies in polarized-bath spin-echo signal. (a)
Simulated quadrature spin-echo signals (Pol. - polarized nuclear bath,
Ref. - unpolarized nuclear bath). We used $B=50$ G, $n=0.1$, and
$P=1$ for the polarized bath case (b) The Q-signals at the first
revival time for various degree of polarization ($B=10$ G, $n=$0.01).
inset: the rotation frequency $\varpi$ as a function of the polarization
$P$ at these conditions. (c,d) The frequency $\varpi$ and the contrast
$C$ as a function of the magnetic field $B$, and $^{13}$C abundance
$n$, for a maximally polarized nuclear spin environment ($P$=1).}
\end{figure}

To conclude, we studied the use of a central spin, realized here by
the NV-center, as a probe for the polarization of a proximal spin
bath (complementary to direct bulk nuclear measurement \cite{FischerPRL2013,AlvarezArXiv2014}).
We demonstrated experimentally that by measuring the full spin-echo
quadrature of an electronic spin one can measure the magnitude and
orientation of a vicinal nuclear spin. We apply this understanding
to the case of a polarized spin bath and found that the electronic
coherence rotates in a characteristic frequency, which is proportional
to the total bath magnetization. Thus, our scheme offers a novel sensing
method for mesoscopic polarized environments. Our sensing method is
insensitive to the nuclei geometrical configuration, in contrast to
the zeeman shift induced by static field measurements. Therefore,
our technique should better apply to nuclear bath polarization in
random environment such as NV-ensembles. Our results emphasize that
the polarization of the central spin surroundings plays a major role
in the spin dynamical behavior.

\bigskip{}

\bigskip{}

\begin{acknowledgments}
The authors thank Chen Avinadav for fruitful discussions and suggestions.
J.R.M acknowledges support from Fondecyt-Conicyt grant No. 1141185,
PIA-Conicyt grant No. ACT1108, and US Air Force grant FA9550-15-1-0113.
\end{acknowledgments}

\bibliographystyle{pnas}
\bibliography{Bibliography,NVreferences}

\begin{thebibliography}{10}
\bibitem{Comment2}
(In general, the electronic $\left|0\right\rangle$ probability after a
  $\phi$-shifted terminating $\left(\pi/2\right)$ pulse is
  $p_{\phi}=\frac{1}{2}\left(1+\mathrm{Re}\left\{ e^{i\phi}S\right\}
  \right)$.).
\bibitem{CommentForBathPOlarizationPaper}
(In ultra-pure diamonds, the lifetime $T_{1}$ of nuclear spins is substantially
  longer than their phase coherence time $T_{2}$. Thus, we assume that at the
  beginning of the sequence the k-th nuclear spin isn't in a pure state, but
  contains some residual polarization).
\end{thebibliography}

\end{document}